\documentclass[lettersize,journal]{IEEEtran}
\IEEEoverridecommandlockouts
\usepackage{cite}
\usepackage{amsmath,amssymb,amsfonts}
\usepackage{algorithm, algorithmic}
\usepackage{graphicx, subcaption}
\usepackage{balance}
\usepackage{textcomp}
\usepackage{xcolor}
\usepackage{booktabs, multirow, multicol}
\usepackage{enumitem}

\usepackage{array,comment} 

\def\BibTeX{{\rm B\kern-.05em{\sc i\kern-.025em b}\kern-.08em
    T\kern-.1667em\lower.7ex\hbox{E}\kern-.125emX}}
\begin{document}

\author{Sefa Kayraklik, Ali Fuat Sahin, Onur Salan, Recep A. Tasci, Recep Vural, Yusuf Islam Tek, Ertugrul Basar, Ibrahim Hokelek, Ali Gorcin, Karim Boutiba, and Adlen Ksentini 
\vspace{-15pt}
        \thanks{Sefa Kayraklik, Ali Fuat Sahin, Onur Salan, Ibrahim Hokelek, and Ali Gorcin are with the HISAR Lab. of TUBITAK-BILGEM, Turkiye; Recep A. Tasci, Yusuf Islam Tek, and Sefa Kayraklik are with Koç University, Turkiye; Recep Vural and Ertugrul Basar were with Koç University, Turkiye; Recep A. Tasci is also with Istanbul Medipol University, Turkiye; Recep Vural and Ertugrul Basar are now with Tampere Wireless Research Center, Tampere University, Finland; Yusuf Islam Tek is also with the Turk Telekom R\&D Department, Turkiye; Ali Gorcin is also with Istanbul Technical University, Turkiye; Karim Boutiba and Adlen Ksentini are with EURECOM, France.}
}

\title{ORIX: Orchestration of RIS with xApps for Smart Wireless Factory Environments}

\markboth{IEEE Wireless Communications}%
{Kayraklık \MakeLowercase{\textit{et al.}}: ORIX: Orchestration of RIS with xApps for SWF}

\maketitle

\begin{abstract} 
The vision of a smart wireless factory (SWF) demands highly flexible, low-latency, and reliable connectivity that goes beyond conventional wireless solutions. Reconfigurable intelligent surface (RIS)-empowered communications, when integrated with the open radio access network (O-RAN) architectures, have emerged as a promising enabler to meet these challenging requirements. This article introduces the methodology for the orchestration of RIS with xApps (ORIX), bringing the RIS technology into the O-RAN ecosystem through xApp-based control for SWF environments. ORIX features three key components: an O-RAN-compliant RIS service model for dynamic configuration, an RIS channel simulator that supports 3GPP indoor factory models with multiple industrial scenarios, and practical RIS optimization strategies with finite-resolution control. Together, these elements provide a realistic end-to-end emulation platform for evaluating RIS placement, control, and performance in SWF environments prior to deployment. The presented case study demonstrates how ORIX enables the evaluation of achievable performance gains, exploration of trade-offs among key RIS design parameters, and identification of deployment strategies that balance system performance with practical implementation constraints. By bridging theoretical advances with industrial feasibility, ORIX lays the groundwork for RIS-assisted O-RAN networks to power next-generation wireless communication in industrial scenarios.

\end{abstract}

\vspace{-5pt}


\section{Introduction}
\label{introduction}
The evolution of next-generation wireless communication systems has introduced key service applications, such as massive machine-type communication (mMTC), ultra-reliable low-latency communication (URLLC), and enhanced mobile broadband (eMBB), which enable the realization of smart wireless factories (SWFs). These advanced services address challenging requirements such as high throughput, mission-critical reliability, and dynamic flexibility in industrial environments \cite{11010844}. Among emerging enablers, the open radio access network (O-RAN) stands out as a promising technology that can support further revolution of the SWF environments for mission-critical requirements handling end-to-end quality of services \cite{10255760}. Through its RAN intelligent controllers (RICs), empowered by open entities and interfaces, the O-RAN framework enables fine-grained and adjustable network management \cite{polese2023understanding}. This openness and programmability facilitate the seamless introduction of new applications, functionalities, and features, while paving the way for the integration of emerging wireless technologies to meet the evolving requirements of SWFs.

The dynamic and highly variable nature of SWF environments necessitates efficient mechanisms for adapting to rapidly changing wireless channel conditions. In this context, reconfigurable intelligent surfaces (RISs) have emerged as a promising candidate for reshaping and controlling the wireless propagation environment, thereby enhancing adaptability and robustness \cite{8796365}. While the 3rd Generation Partnership Project (3GPP) Release 18 recently introduced network-controlled repeaters (NCRs), which require active hardware like power amplifiers and radio frequency chains that amplify both the signal and thermal noise, RISs utilize passive or semi-passive components such as diodes to provide noise-free reflection.
However, RISs also face several limitations, including limited reflection efficiency, a lack of standardized control interfaces, and challenges in channel estimation. The adoption of NCRs by 3GPP can be attributed to their technological maturity, compatibility with existing network architectures, and predictable performance, whereas the RIS technology is still under active research and lacks a unified standardization framework \cite{10467188}.
To this end, key RIS adoption challenges in next-generation wireless networks need to be addressed to provide seamless integration through realistic modeling and control mechanisms within a standardized architecture, while ensuring efficient RIS optimization under practical constraints.

Prior to the deployment of an RIS in real-world SWF scenarios, critical design considerations, including the placement and specifications of the RIS, such as the number of reflecting elements, quantization levels of the phase shifts, and the operating frequency, must be carefully addressed. Hence, accurate simulation models for RIS-assisted channels in factory environments are essential to determine key parameters and validate the optimization algorithms in advance \cite{11106752}. Ultimately, a fundamental challenge lies in effectively modeling and orchestrating the RIS operation within the O-RAN framework to satisfy the critical requirements of SWFs.

Recent studies have begun exploring the integration of the RIS into the O-RAN architecture. For example, the performance of the RIS-aided communication under different network management policies controlled by xApps is evaluated on an emulated O-RAN platform for the eMBB and URLLC network slices \cite{tsampazi2025ris}. An experimental prototype also showcases the RIS deployment within the O-RAN framework, where customized xApps are developed to monitor users’ channel quality and execute multi-user RIS optimization in real time \cite{sahin2025ris}. In addition, a Golang-based RIS simulator (GoSimRIS) is introduced to enable real-time RIS control through an O-RAN RIS service model, directly interfacing with the RIC \cite{el2025neural}.

\begin{figure*}[t]
    \centering
    \includegraphics[width=1\textwidth]{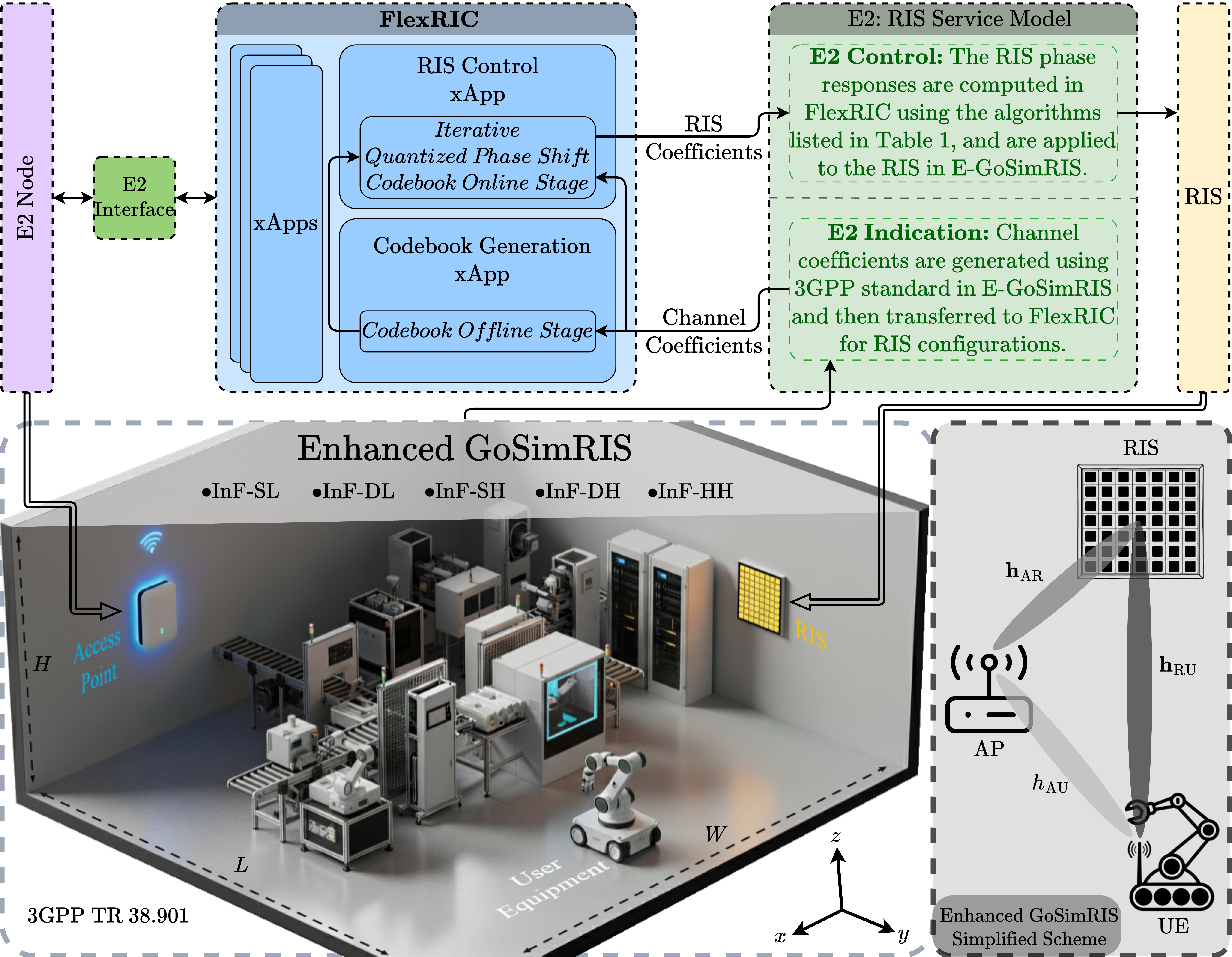}
    \caption{ORIX system architecture of the O-RAN integrated RIS-assisted indoor SWF environments} 
    \label{fig:ORIX}\vspace{-10pt}
\end{figure*}

Motivated by the potential of O-RAN-based RIS-assisted communications in enabling SWF environments, this article introduces the orchestration of RIS with xApps (ORIX) framework for indoor industrial scenarios. Unlike generic integration efforts, ORIX seamlessly integrates the RIS technology into the O-RAN architecture to address the dynamic connectivity and control requirements of SWFs. To bridge the gap between O-RAN protocols and industrial physical environments for RISs, ORIX features three key components: an O-RAN-compliant open interface service model specialized for RIS control, enabling fast response to rapid fading profiles of industrial settings; an RIS simulator enhanced with the 3GPP-based indoor factory (InF) channel models with realistic RIS hardware constraints, enclosing specific clutter densities and machinery topologies; and a set of practical RIS optimization strategies under real-world deployment considerations. ORIX helps address key challenges in RIS adoption by improving modeling accuracy and enabling practical and efficient control in realistic deployment scenarios. With ORIX, realistic factory environments can be emulated to explore RIS placement and orchestration strategies and assess their performance impact within an O-RAN ecosystem well before physical deployment.

\section{Enabling Technologies for Smart Wireless Factories} \vspace{-5pt}
\label{sectionII}
SWFs are modern industrial sites that leverage machine-type, reliable, and low-latency wireless connectivity to boost efficiency, improve safety, and support rapid reconfiguration. As industrial demands continue to rise, driven by mass customization and shorter product cycles, conventional wireless architectures struggle to satisfy the challenging requirements of high throughput, strict latency, and flexibility targets. To overcome these limitations, factories must transition toward intelligent wireless frameworks that deliver the robustness, scalability, and resilience required by modern manufacturing. In this context, integrating the O-RAN framework, which brings openness, programmability, and intelligent control to factory connectivity, with the RIS technology, which can dynamically shape the wireless environment to ensure reliable and efficient communication, forms a powerful combination. The O-RAN framework provides software-driven intelligence for adaptive network management, while the RIS extends this adaptability into the physical propagation domain. The O-RAN-based RIS-assisted SWF environment is illustrated in Fig. \ref{fig:ORIX}, which is described in Section \ref{sec:arch}. By adopting such adaptable frameworks, SWFs gain greater operational flexibility, higher throughput, and a more resilient environment. 

\subsection{Communication and Control Requirements} 
Despite the advantages offered by SWFs, their real-world deployment introduces strict communication and control requirements. A primary objective is to maintain reliable, robust, and flexible wireless networks that interconnect diverse factory components such as machines, conveyor belts, and cranes. The dense deployment of devices in an indoor environment increases the demands on both the control and data planes. 

From the communication perspective, the indoor factory environment is a challenging medium, where line-of-sight (LoS) is rarely available due to the constant movement of machinery. Rich multipath and shadowing further degrade the signal quality, while reflective metal walls and equipment worsen the signal propagation. Beyond indoor factory challenges, industrial applications impose even stricter conditions. The safety and mission-critical operations require communication links capable of providing ultra-low-latency and extremely low packet-error rates since even a single corrupted packet could trigger hazardous outcomes. 

The control requirements of SWF environments are similarly challenging due to the potentially hundreds of wireless nodes within the network, with the top priority of seamless and scalable network optimization. Even minor latency spikes or packet losses can cascade into quality faults or safety risks. To address these challenges, control loops must operate within strict timing constraints without overloading the medium, controllers must be delay-aware, and redundant loops may be incorporated to preserve stability during transient outages.

\subsection{Overview of O-RAN and RIS} 
O-RAN reshapes the conventional, vendor-dependent RAN stack as a disaggregated, cloud-native platform built on open entities and interfaces \cite{polese2023understanding}. By splitting the access network into radio units (O-RU), centralized units (O-CU), and distributed units (O-DU) and layering an intelligent control plane managed by RICs, O-RAN enables data-driven network management and supports cost-effective deployment. The RIC framework leverages rich RAN telemetry and is logically partitioned by control periodicity. The non-real-time RIC (non-RT RIC) executes long-standing tasks ($\geq 1$s) through rApps, such as artificial intelligence (AI) and machine learning (ML)-based optimization, policy management, and network analytics, while the near-real-time RIC (near-RT RIC) governs short-term tasks ($10$ms-$1$s) through xApps, such as interference management, handover optimization, and radio resource scheduling. For general readers, it is helpful to view the RIC as an operating system for the radio network. In this analogy, rApps/xApps function like specialized software applications installed on a smartphone. This modular approach enables operators to dynamically install new network capabilities without requiring changes to existing hardware. Open interfaces such as A1, O1, and E2 expose the required telemetry data, allowing custom logic to be implemented in both rApps and xApps. Consequently, the openness of O-RAN introduces vendor diversity, reduces costs, and accelerates RAN innovation and leading the way to affordable, vendor-agnostic private networks, which are essential for deploying SWFs in industrial settings. 

RIS-empowered communications have emerged as a promising enabler for next-generation wireless communication networks, introducing a paradigm shift in which the propagation environment can be reconfigured \cite{8796365}. Conceptually, an RIS acts like a smart mirror for radio waves. Unlike a metal surface that reflects signals in a fixed direction like a mirror reflecting light, an RIS consisting of a large number of programmable elements can be electronically adjusted to steer the reflected signal toward a specific direction. By coating walls, ceilings, and other large surfaces with passive metasurface tiles whose reflection coefficients can be dynamically adjusted in real time, an RIS can alter the phase and/or amplitude of incident signals, thereby converting hostile multipath into a controllable resource. This capability boosts link budget and supports energy-efficient transmitter designs. Moreover, RISs can also extend the coverage of millimeter wave links, enhance the physical-layer security, and improve energy efficiency by enabling low-power beam steering. Strategically deployed in the SWF environment, RISs can create a virtual LoS path when machinery blocks the direct paths and reinforce existing LoS links to improve reliability, latency, and coverage. Accordingly, the RIS technology is widely viewed as one of the cornerstones of next-generation communication and a practical enabler for a resilient, flexible communication infrastructure for SWFs. 

\subsection{Integration Challenges}
The integration of RISs and O-RAN into an SWF poses several practical obstacles despite their complementary advantages. First, the highly dynamic propagation conditions in industrial settings raise difficulties in pre-calibrating RIS layouts through conventional methods, often necessitating comprehensive channel measurements. Second, many commercial private network deployments remain vendor-locked, limiting the interoperability and flexibility that O-RAN promises. Lastly, integrating the RIS with the O-RAN framework is still an emerging area in which integration protocols are not yet standardized. These challenges highlight the need for a unified framework that can accurately capture factory-specific channel characteristics, enable realistic RIS simulations, minimize vendor dependence, and guide the development of robust standardization efforts. Such a framework is critical to unlocking the full potential of RIS and O-RAN integration in SWF environments.

\section{System Architecture: Orchestration of RIS with xApps for Indoor SWF Scenarios} \label{sec:arch}
Building on the challenges outlined earlier, the ORIX framework integrates RISs into O-RAN-based indoor SWF networks, whose overall system architecture is depicted in Fig. \ref{fig:ORIX}, where the O-RAN orchestration framework, the emulated SWF environment, and the corresponding simplified simulation scheme are employed on the top and the bottom-left, and bottom-right, respectively. In the physical emulations, the RIS channel simulator provided by enhanced GoSimRIS (E-GoSimRIS) models standard-compliant 3GPP InF propagation channels between the access point (AP), the user equipment (UE), and the RIS. The RIS control loop begins with E2 indications conveying channel coefficients upward to the near-RT RIC. After the dedicated xApps execute the selected RIS phase optimization approach, the loop is closed by E2 control messages, which send the calculated RIS coefficients back down to dynamically reconfigure the emulated RIS. Namely, ORIX serves as an end-to-end simulation and emulation platform designed to support the development of xApp-based controllers for open RIS platforms and their seamless integration into next-generation wireless communication systems with the O-RAN framework. 

Before network-controlled RISs can be deployed in a real-world O-RAN network, control algorithms must be carefully validated, tuned, and benchmarked to ensure their ability to meet industrial performance requirements. Conducting such evaluations directly in a live factory setting can be costly and disruptive. ORIX addresses this gap by providing a realistic end-to-end emulator of RIS-enabled, O-RAN-based private networks for indoor SWF scenarios. ORIX enables the modeling of indoor factory environments, experimentation with RIS placement and control strategies, and quantification of their performance impact within the O-RAN architecture well before physical deployment.

\subsection{E2 Service Model Extension for RIS Control} 
The first key component of ORIX is the E2 node. In O-RAN, E2 nodes represent RAN elements that interface with the near-RT RIC over the E2 interface, enabling real-time reporting and resource control \cite{polese2023understanding}. Typical E2 nodes include O-CU, O-DU, O-RAN-compliant evolved NodeBs, and next-generation NodeBs (gNBs). In ORIX, a \textit{monolithic} structure of the co-located CU and DU is employed with an E2 agent that provides the logical behavior of an E2 node, establishes the first element as an AP. The second element is the E2 interface, which serves as the communication and control backbone between E2 nodes and near-RT RIC \cite{polese2023understanding}. The E2 interface is structured around two complementary protocols, which are the E2 application protocol (E2AP) and the E2 service model (E2SM). The E2AP is a control protocol over the E2 interface with the aim of governing the communication between near-RT RIC and E2 nodes. Similarly, E2SM is the content structure that defines the specific telemetry and control logic in E2AP messages. In other words, E2AP defines \textit{how} near-RT RIC communicates with E2 nodes while E2SM defines \textit{what} is exchanged. Together, E2AP and E2SM provide the dual foundation for real-time monitoring and smooth control in ORIX.

While the O-RAN Alliance standardizes E2AP and a set of baseline E2SMs, custom E2SMs can be designed to support new applications. In ORIX, the E2 RIS service model (E2SM-RIS) has been implemented as a standardized service model specially designed to integrate the RIS into the O-RAN control loop \cite{el2025neural}. By designing the RIS as a standardized E2 node, ORIX aims to establish E2SM-RIS as the critical interface that allows hardware and software components from heterogeneous vendors to remain interchangeable, provided they adhere to a unified messaging structure. Consequently, this model serves as a foundational reference for the development of more sophisticated RIS control methods and future formalization within O-RAN specifications.

The E2SM-RIS model in Fig. \ref{fig:ORIX} defines the information exchange between near-RT RIC and an RIS to enable dynamic configuration of the reflecting elements within the O-RAN network. Two classes of messages are employed in E2SM-RIS: E2 indication messages from the E2 node to the near-RT RIC and E2 control messages from near-RT RIC to the E2 node. E2 indication messages serve as the primary telemetry mechanism, reporting channel-state information between the AP-RIS, RIS-UE, and AP-UE channels together with structural RIS parameters such as RIS dimensions, the number of RIS elements, and element aperture. Consequently, ORIX can continuously monitor the link quality. On the other hand, E2 control messages convey high-level commands, such as desired beam orientations or per-element phase shift coefficients, obtained from xApps decisions, to the E2 node.

\subsection{RIC Framework in ORIX}
For the near-RT RIC framework, ORIX adopts FlexRIC \cite{schmidt2021flexric}, an open-source implementation that offers an extensible framework for developing and deploying custom xApps while complying with O-RAN specifications. FlexRIC supports the O-RAN-defined E2 interface, including E2AP and multiple standardized E2SMs, thereby ensuring interoperability with other O-RAN entities. Furthermore, O-RAN standardization provides inherent architectural safeguards for signaling efficiency and interface security. For instance, O-RAN standards leverage mandatory protocols to secure the E2 interface and support various compression techniques to improve signaling efficiency, ensuring the integrity and feasibility of the RIS control loop in mission-critical industrial settings \cite{polese2023understanding}.

By leveraging the O-RAN E2 feedback loop to maintain reliability, ORIX has the ability to adapt to time-varying channel conditions, such as those caused by moving machinery or temporal blockages. The process begins with the RIS control xApp continuously monitoring the channel quality by subscribing to periodic E2 indication reports, which contain channel coefficients. When these reported metrics drop, indicating a sudden blockage or significant channel degradation, the xApp immediately triggers its optimization routine to calculate a new phase configuration for the RIS. As a near-RT RIC application operating within a $10$ms to $1$s control loop, ORIX can detect these environmental shifts and send updated phase configurations via E2 control messages, effectively redirecting the signal path before the connection is lost.

Furthermore, adhering to O-RAN standards enables ORIX to seamlessly integrate non-RT RIC capabilities and learning-based applications into the RIS orchestration loop. By following the standard O-RAN AI/ML workflow, computationally intensive training tasks are decoupled from the real-time control loop and offloaded to the non-RT RIC, where latency requirements are more flexible \cite{polese2023understanding}. This modular structure ensures that the high resource demands of training do not interfere with the mission-critical, low-latency connectivity required on the factory floor. ORIX can also support latency-critical applications by deploying lightweight, pretrained AI/ML models as modular xApps, provided they are optimized to operate within the near-RT control window.

\subsection{RIS Simulator of ORIX for Indoor SWF}
The RIS channel simulator within ORIX is provided by E-GoSimRIS, the third generation of SimRIS simulators, whose first version is an open-source MATLAB-based channel simulator for RIS-assisted millimeter wave (mmWave) channels that follows the 3GPP channel models for both indoor and outdoor scenarios \cite{basar2020simris}. The second version, GoSimRIS, is a Go-based RIS simulator \cite{el2025neural}, which is designed for real-time communication with remote entities while meeting low-latency requirements. Building upon these, E-GoSimRIS introduces support for the 3GPP InF channels for mmWave links \cite{etsi2020138} and provides five deployment scenarios for SWF environments, capturing different conditions such as LoS availability, dominant reflections, and delay spreads. Those scenarios are comprehensively explained in Section \ref{sec:risdepstr}.

To meet strict near-RT RIC latency constraints, ORIX leverages the high-concurrency capabilities of the Go language. Additionally, ORIX can further be architected to support advanced queuing logic and multi-threaded xApp architectures, ensuring that the $10$ms-$1$s time-frame is maintained even when the system scales to manage a large number of devices or multiple RIS panels. With E-GoSimRIS, ORIX can emulate realistic industrial propagation environments, enabling the detailed evaluation of RIS-assisted O-RAN deployments in SWF networks. Therefore, E-GoSimRIS is a critical element of the ORIX architecture because conducting empirical validation in live factory environments is often prohibitively costly and operationally disruptive. By providing a realistic end-to-end emulation platform, it allows for the experimentation of RIS placement and control strategies before any physical deployment occurs, ensuring that industrial performance requirements are met safely and efficiently. Furthermore, by adhering to the 3GPP InF scenarios, ORIX incorporates statistical representations of temporal dynamics, such as Doppler spread, time-varying path gains, and blockage modeling. This ensures that the simulation captures the statistical essence of rapidly fluctuating factory conditions.

\begin{table}[t!]
\caption{Practical RIS phase optimization methods}
\centering
\begin{tabular}{|>{\centering\arraybackslash}p{0.15\linewidth}|>{\centering\arraybackslash}p{0.75\linewidth}|}
\hline
\textbf{Method} & \textbf{Steps}  \\
\hline
\vspace{0.65cm}
\textbf{Iterative \cite{kayraklik2024indoor}} & 
\begin{enumerate}
    \item Initialize all RIS elements.
    \item Sweep each element once. For each element;
    \begin{itemize}[label=-]
        \item Test all possible phase levels.
        \item Keep the phase that yields the highest rate.
    \end{itemize} 
    \item After one full sweep, optimization is finalized.
\end{enumerate}  \\
\hline
\vspace{0.12cm}
\textbf{Quantized Phase Shift}  &
\begin{enumerate}
    \item Compute the ideal phase, aligning the cascaded and direct paths.
    \item Map the phase to the nearest discrete level.
\end{enumerate}  \\
\hline
\vspace{1.35cm}
\textbf{Codebook} &
\begin{itemize}
    \item Offline Stage:
    \begin{enumerate}
        \item Sweep many channel realizations.
        \item Store the phase vector that maximizes the rate for each codebook position.
    \end{enumerate}
    \item Online Stage:
    \begin{enumerate}
        \item Read the current channel estimate.
        \item Sweep each codebook once. Choose the one that provides the best rate.
        \item Apply the chosen codebook to the RIS.
    \end{enumerate}
\end{itemize}  \\
\hline
\end{tabular}
\label{tab:ris_methods}
\end{table}

\subsection{RIS Optimization in ORIX} \label{sec:opti}
In the ORIX framework, the primary optimization objective is to maximize the achievable data rate as described in \cite[eq. (2)]{el2025neural}. The RIS facilitates this by adjusting the phase shifts of incident signals to ensure the reflected signals steer towards the UE in phase with the signals from the direct path. These constructive paths effectively boost the received signal power, creating a virtual LoS link that significantly enhances the achievable data rate and throughput, even when the direct path is blocked or weak. In practice, the RIS elements support a finite set of phase shift adjustments. ORIX enforces this finite-resolution control to emulate practical RIS hardware. The optimization problem is therefore to determine the RIS phase shift adjustment that maximizes the data rate under the discrete phase constraint. Additionally, ORIX leverages the near-RT RIC control loop to enable dynamic reconfigurations, providing a significant advantage over static RIS configurations. While static RIS deployments are optimized for fixed coverage and cannot adapt to environmental shifts, ORIX can adapt to adverse channel conditions such as sudden blockages or temporal dynamics caused by moving machinery or robotic arms. This adaptive capability is essential for ensuring low-latency industrial applications.

ORIX provides three complementary RIS optimization methods, whose steps are summarized in Table \ref{tab:ris_methods}. The first is the iterative method \cite{kayraklik2024indoor}, in which the RIS elements are optimized sequentially. For each element, all possible discrete phase shifts are evaluated to observe data rate improvement while keeping the other elements fixed, and the phase with the highest rate is selected. This method can achieve near-optimal performance with computational complexity proportional to the number of RIS elements and the phase shift resolution without an exhaustive search, which makes it suitable for practical applications. The second approach is the quantized phase shift method with computational complexity proportional to the number of RIS elements, where the optimal continuous phase aligning the cascaded and direct paths is computed and then mapped to the nearest discrete level. Finally, the codebook method operates with offline/online steps. In the offline stage, a library of phase shift configurations is constructed by maximizing the achievable data rate over representative channel realizations. During the online stage, the controller with computational complexity proportional to the number of predefined codebooks selects the most suitable codebook entry based on current channel estimates, requiring only a single-pass evaluation. 

\begin{table*}[t]
\caption{Indoor factory (InF) deployment scenarios and parameters}
\centering
\renewcommand{\arraystretch}{1.4}
\setlength{\tabcolsep}{5pt}
\begin{tabular}{|>{\centering\arraybackslash}p{1.4cm}
                |>{\centering\arraybackslash}p{2cm}
                |>{\centering\arraybackslash}p{2.25cm}
                |>{\centering\arraybackslash}p{2.25cm}
                |>{\centering\arraybackslash}p{2.25cm}
                |>{\centering\arraybackslash}p{2.25cm}
                |>{\centering\arraybackslash}p{2.25cm}|}
\hline
\multicolumn{2}{|c|}{\textbf{Parameter Category}} & \textbf{InF-SL} & \textbf{InF-DL} & \textbf{InF-SH} & \textbf{InF-DH} & \textbf{InF-HH} \\
\hline
\multirow{4}{1.4cm}{
\parbox[c][2cm][c]{1.45cm}{\centering
Layout
}
} & Room size ($A_r$) & \multicolumn{5}{c|}{Rectangular: $20$-$160{,}000$ m\textsuperscript{2}} \\
\cline{2-7}
& Ceiling height ($h_{\text{ceil}}$) & \raisebox{-5pt}{\parbox{2cm}{\centering $5-25$ m}}  & \raisebox{-5pt}{\parbox{2.25cm}{\centering $5-15$ m}} & \raisebox{-5pt}{\parbox{2.25cm}{\centering $5-25$ m}} & \raisebox{-5pt}{\parbox{2.25cm}{\centering $5-15$ m}} & \raisebox{-5pt}{\parbox{2.25cm}{\centering $5-25$ m}} \\
\cline{2-7}
& Effective clutter height ($h_c$) & \multicolumn{5}{c|}{\raisebox{-5pt}{\parbox{5.25cm}{\centering Less than $h_\text{ceil},\ 0-10$ m}}} \\
\cline{2-7}
& External wall and ceiling type & \multicolumn{5}{c|}{\raisebox{-5pt}{\parbox{10cm}{\centering Concrete or metal walls and ceiling with metal-coated windows}}} \\
\hline
\multicolumn{2}{|c|}{\raisebox{-12pt}{\centering Clutter Type}} & \raisebox{-12pt}{\parbox{2.25cm}{\centering Big machineries with regular metallic surface}} & \raisebox{-12pt}{\parbox{2.25cm}{\centering Small to medium metallic machinery with irregular structure}}\vspace{3pt} & \raisebox{-12pt}{\parbox{2.25cm}{\centering Same as InF-SL}} & \raisebox{-12pt}{\parbox{2.25cm}{\centering Same as InF-DL}} & \raisebox{-12pt}{\parbox{2.25cm}{\centering Any}} \\
\hline
\multicolumn{2}{|c|}{Typical clutter size ($d_{\text{clutter}}$)} & $10$ m & $2$ m & $10$ m & $2$ m & Any \\
\hline
\multicolumn{2}{|c|}{\raisebox{-5pt}{\parbox{3.25cm}{\centering Clutter density ($r$)}}} & Low clutter density ($< 40\%$) & High clutter density ($\geq 40\%$) & Low clutter density ($< 40\%$) & High clutter density ($\geq 40\%$) & \raisebox{-5pt}{\parbox{2.25cm}{\centering Any}} \\
\hline
\multicolumn{2}{|c|}{AP antenna height ($h_{\text{AP}}$)} & \multicolumn{2}{c|}{Clutter-embedded} & \multicolumn{3}{c|}{Above clutter}  \\
\hline
\multirow{2}{*}{UE location} & LOS/NLOS & \multicolumn{4}{c|}{LOS and NLOS} & 100\% LOS \\
\cline{2-7}
& Height ($h_{\text{UE}}$) & \multicolumn{4}{c|}{Clutter-embedded} & Above clutter \\
\hline
\end{tabular}
\label{table:deployment_parameters}
\end{table*}

\section{Implementation Considerations and Practical Challenges}
\label{sectionIV}
To ensure accurate assessment of the RIS behavior in industrial settings, the simulator is extended with a 3GPP-compliant indoor factory model, as defined in TR 38.901. This enhancement provides a realistic propagation environment, including multipath and blockage effects typical in smart manufacturing facilities.

\subsection{Simulation Environment}\label{sec:risdepstr}
The deployment of an RIS in InF environments requires careful consideration of the diverse propagation conditions and physical constraints typical of industrial facilities. Following the 3GPP TR 38.901 specification \cite{etsi2020138}, the InF scenario is modeled as a rectangular industrial hall, characterized by large dimensions ($20-160,000$ m$^{2}$), ceiling heights between $5-25$ m, and clutter elements such as metallic machinery, conveyor systems, and storage racks. These environments inherently exhibit rich multipath propagation, shadowing, and frequent non-LoS conditions, strongly motivating the usage of an RIS to enable controllable and reliable communication links.

In order to capture the heterogeneity of factory environments, five scenarios are defined according to the clutter density and the heights of AP/UE \cite{etsi2020138}: 
\begin{itemize}
    \item Sparse clutter, Low AP/UE (InF-SL),
    \item Dense clutter, Low AP/UE (InF-DL),
    \item Sparse clutter, High AP/UE (InF-SH),
    \item Dense clutter, High AP/UE (InF-DH), and 
    \item High AP and High UE (InF-HH).
\end{itemize}
Specifically, the classification depends on two main parameters: clutter density, which ranges from $<40\%$ for sparse layouts to $\geq40\%$ for dense shop floors, and relative antenna heights with respect to the average clutter level. For example, the InF-SL and InF-DL model cases, where both the AP and the UE are embedded within machinery, while InF-HH assumes elevated transceivers above clutter, resulting in almost guaranteed LoS conditions. Table \ref{table:deployment_parameters} summarizes the geometric and material properties of the InF environment, including clutter size, wall composition, and antenna heights.

In RIS-assisted InF environments, metasurfaces are strategically deployed on walls or ceilings to compensate for blockage, enhance coverage, and enable programmable reflections. As illustrated in Fig. \ref{fig:ORIX}, the utilization of an RIS introduces additional wireless channels ($\mathbf{h}_\text{AR}$ and $\mathbf{h}_\text{RU}$) in addition to the LoS channel ($h_\text{AU}$). Accurate modeling of these links is essential: while $\mathbf{h}_\text{AR}$ and $\mathbf{h}_\text{RU}$ are often assumed LoS due to the placement of RISs at higher elevations in the factory environments, $h_\text{AU}$ remains subject to probabilistic LoS conditions determined by clutter density and distribution. This dual-link structure is explicitly accounted for in the E-GoSimRIS simulator, which generates channel coefficients for each sub-scenario by incorporating large-scale parameters (delay spread, angular spreads, shadow fading) and small-scale multipath effects in compliance with TR 38.901 \cite{etsi2020138}.

\subsection{Practical RIS Models for Optimization Algorithms}
To account for practical implementation limitations of the RIS, E-GoSimRIS considers and models phase shifts that are dependent on bit resolution and reflection amplitudes that vary with phase.

\begin{figure*}[t]
\centering
\includegraphics[width=0.99\linewidth]{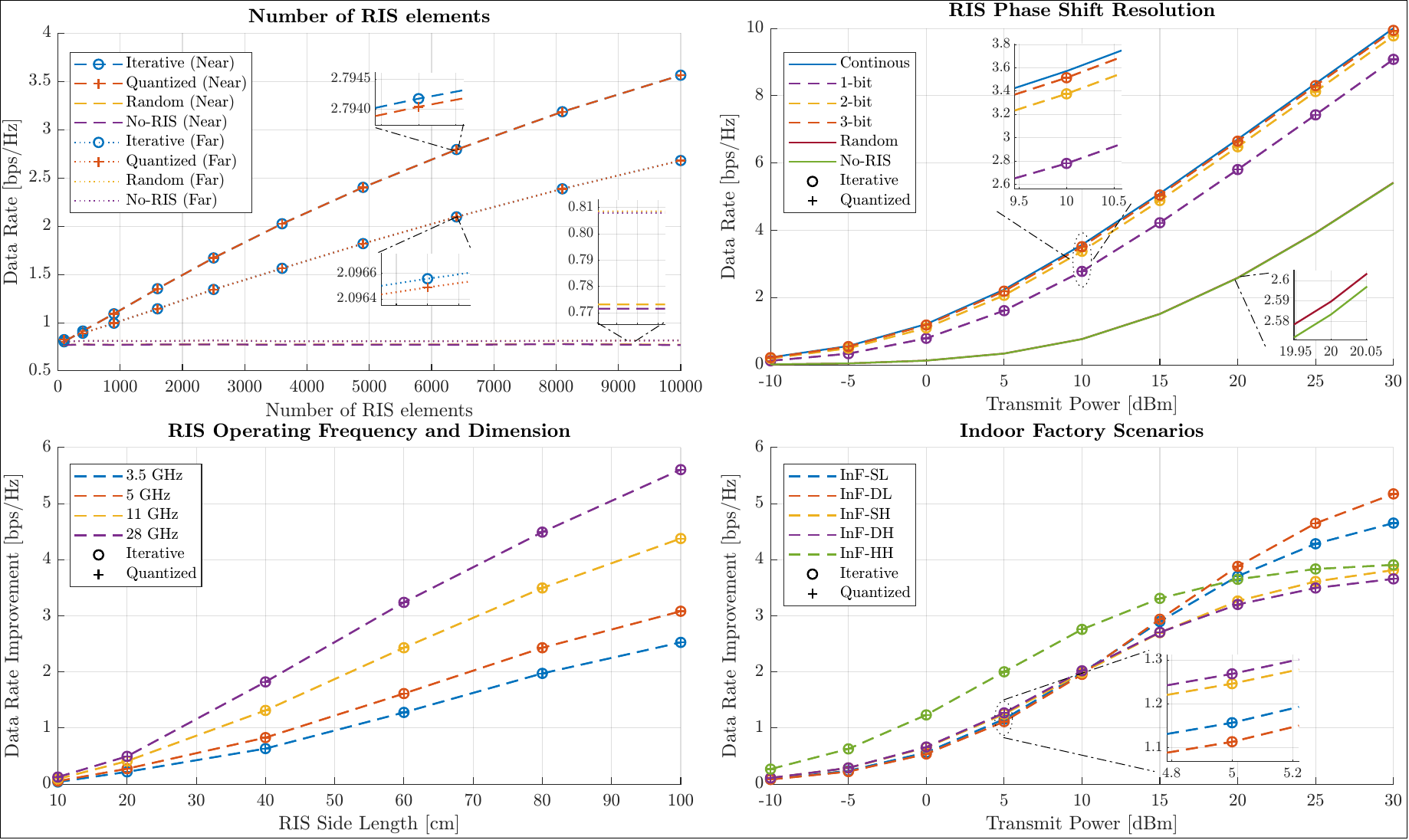}
\caption{Assessment of the RIS design parameters for SWF environments}
\label{fig:simScenario}
\end{figure*}

\subsubsection{Phase Shift Model}
Practical RIS hardware cannot achieve infinite precision; instead, the phase shift of each reflecting element is controlled using a finite number of bits. This results in a discrete set of possible phase shifts equally spaced over the full circle ($360$ degrees), rather than a continuous range. To serve as a theoretical benchmark, a continuous phase shift model is also included, representing an ideal scenario with infinite resolution, even though such fine-grained control remains impractical. Furthermore, the performance gain of the continuous phase shift model is not significantly better than that of the two-bit phase shift model, highlighting the diminishing returns of higher bit resolutions in practice. The control delay for RIS reconfiguration is neglected because the time to adjust the phase shift is much shorter than the average industrial channel coherence time.

\subsubsection{Reflection Amplitude Model}
In much of the existing RIS literature, the reflections from RIS elements are often assumed to be lossless. However, this assumption does not reflect the characteristics of practical RIS hardware, where amplitude degradation is commonly observed. In the implemented algorithms, this limitation is explicitly accounted for by adopting the practical model provided in \cite[eq. (5)]{abeywickrama2020intelligent}. This model captures the amplitude loss observed when the phase shift deviates from certain optimal points. The model parameters are typically determined by the specific hardware design and can be estimated from empirical measurements and curve-fitting techniques. ORIX adapts the RIS model parameters as in \cite{jung2021optimality}.
In addition, the mutual coupling between RIS elements is neglected, as its effect becomes negligible when the inter-element spacing is at least half the wavelength, which is satisfied in the ORIX. 

\subsection{Case Study: RIS Deployment in an SWF Environment}
To illustrate the potential of RIS-assisted connectivity in SWF environments and evaluate the various RIS deployment specifications, a case study is conducted using the 3GPP-based indoor factory channel models. The objective is to assess the performance of practical RIS optimization methods and different RIS design parameters, including the number of reflecting elements, phase shift resolution, operating frequency, and physical dimensions, while also capturing the impact of varying factory propagation conditions. To keep the system simple and reduce signaling overhead in low-latency industrial settings, a single-RIS setup with scalable reflecting elements instead of multiple surfaces is considered. Although trends across the transmit power, UE location, and RIS element number are studied, the ORIX framework is flexible and can be easily customized with different frequencies, various factory layouts, clutter levels, antenna heights, and hardware parameters to represent a wide range of industrial environments. Furthermore, ORIX’s modular architecture supports extension to multi-user and mobile scenarios through specialized xApps.

For comprehensive analysis, Monte Carlo simulations with $10^5$ realizations are carried out in MATLAB environment for a representative factory hall of $75$ m in length, $50$ m in width, and $10$ m in height. An AP operating at $28$ GHz with $10$ dBm transmit power is positioned at $(30,0,8)$ m, while a one-bit RIS with $6400$ reflecting elements is mounted at $(75,30,6)$ m. Two UE locations are considered: a near case at $(72,32,1.5)$ m and a far case at $(62,22,1.5)$ m, with a received noise level of $-88$ dBm. The SWF environment is modeled using the InF-DH channel with a clutter density of $0.6$ and an effective clutter height of $2$ m.

\begin{figure}[t]
\centering
\includegraphics[width=0.99\linewidth]{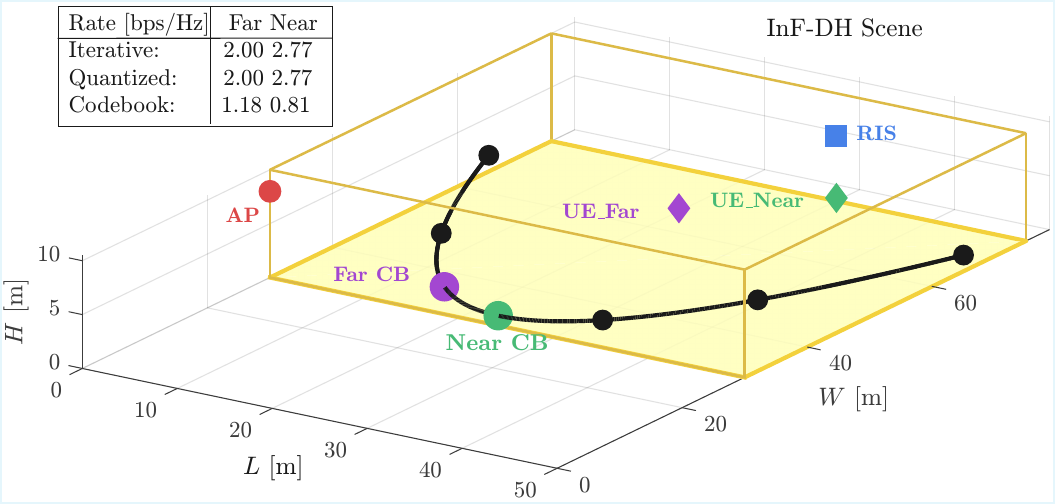}
\caption{InF-DH scene for the RIS-assisted SWF environment}
\label{fig:simScene}
\end{figure}

The practical RIS optimization methods, provided in Section \ref{sec:opti}, are evaluated for near and far UE locations, as illustrated in Fig. \ref{fig:simScene}. For the codebook-based approach, a library of seven RIS configurations, whose positions are denoted as black points, is prepared in the offline stage. During the online stage, the configuration that performs the best is selected. Computer simulation results reveal that, for both user positions, the iterative and quantized methods achieve very similar and higher performance compared to the codebook method. Although the codebook approach underperforms in terms of achievable rate, it offers practical advantages: it eliminates the need for continuous feedback required by the iterative method and avoids the precise channel estimation demanded by the quantized method. However, its performance is highly sensitive to the user’s position, making it best suited for scenarios with stable or predictable user placements.

The design of an RIS for SWF environments requires careful tuning of key parameters such as the number of reflecting elements, phase shift resolution, operating frequency, and physical dimensions. To guide these choices prior to real-world deployment, simulations are carried out, with representative results shown in Fig. \ref{fig:simScenario}. As expected, increasing the number of RIS elements improves the data rate, with the near-UE scenario perceiving more improvement from the RIS. Hence, determining the RIS element number depends not only on the target system performance objectives but also on where the RIS will improve the quality of service. Phase shift resolution is another critical factor: higher resolution directly enhances performance, and with as few as three bits, the results nearly converge to those of continuous phase control. This highlights a practical trade-off between hardware complexity and the desired performance gain. Finally, RIS dimensions are inherently tied to the operating frequency, since the element spacing is mostly set to half the wavelength to mitigate mutual coupling. At higher frequencies, the same aperture accommodates a greater number of reflecting elements, enabling higher performance gains.

All InF scenarios in Table \ref{table:deployment_parameters} are analyzed for the RIS-assisted wireless communication in SWF environments, and the corresponding results are given in Fig. \ref{fig:simScenario}. As expected, the InF-HH scenario exhibits the highest performance gains for low transmit power levels due to the presence of the LoS link. However, as the transmit power increases, the rate improvement converges as the LoS path dominates the channel. Similarly, the InF-SH and InF-DH scenarios exhibit slightly higher performance gains for low transmit powers due to the AP being embedded above the clutter. On the contrary, both InF-SL and InF-DL exhibit modest gains at lower power levels but surpass the other scenarios for higher power levels, owing to the rich presence of multipath links that enable the RIS to deliver larger data rate improvements. 

\section{Conclusion and Future Directions}
This article has presented ORIX, a new framework that integrates RISs into the O-RAN ecosystem for SWF environments. By combining an O-RAN-compliant RIS control interface, a 3GPP-based InF channels of the RIS simulator extended for industrial scenarios, and practical optimization strategies, ORIX enables realistic modeling, experimentation, and evaluation of RIS-assisted SWF deployments. The framework bridges the gap between theoretical advancements and industrial practice by providing a playground to test RIS designs, placement strategies, and orchestration mechanisms within an open and interoperable O-RAN architecture. Furthermore, RIS deployment offers significant performance gains in dense clutter scenarios (InF-DL/DH) by mitigating obstacles through virtual LoS, whereas the improvements are marginal in elevated transceivers (InF-HH), suggesting operators should prioritize the RIS for floor-level assembly lines rather than overhead systems. While higher RIS elements yield greater performance improvements, the RIS design is constrained by the frequency-dependent geometric aperture, requiring a careful balance between beamforming gains and physical installation feasibility. Regarding operational trade-offs, practical RIS control strategies must align with the factory's mobility profile: the codebook-based approach provides low-latency coverage in dynamic environments, while iterative optimization delivers superior precision in static, high-throughput zones. Together, these efforts will help realize the full potential of RIS-aided O-RAN systems in delivering flexible, reliable, and high-performance wireless connectivity for next-generation industrial environments.

Several challenges and open research issues exist for RIS-assisted indoor factory scenarios. Among the potential research directions, the development of closed-loop optimization frameworks in which RIS configurations can be dynamically adapted to evolving factory conditions is a key area of focus. The integration of AI-driven xApps offers the potential to enable continuous monitoring, learning, and optimization. Another promising avenue is the integration of RISs with next-generation wireless technologies such as integrated sensing and communication (ISAC), cell-free multiple-input multiple-output (MIMO), and digital twin technologies. This integration can enhance communication reliability, provide advanced sensing capabilities, and enable virtual testing and deployment planning. However, the lack of standardized interfaces, interoperability frameworks, and accurate channel models for dynamic industrial environments still makes real implementation difficult. Furthermore, validating such an end-to-end simulation environment on physical hardware remains a promising direction for future work. To overcome these challenges, researchers, industry, and standardization groups need to work together so that RIS can move from a research idea to a reliable tool for smart manufacturing. 

\section{Acknowledgment}
ORIX Project has been selected for funding in the first open call of the 6G-BRICKS project. We thank the 6G-BRICKS Project of the European Union’s Horizon Program under Grant Agreement No. 101096954.

\bibliographystyle{IEEEtran}
\bibliography{main.bib}

\vspace{-40pt}
\begin{IEEEbiographynophoto}{SEFA KAYRAKLIK} 
(Graduate Student Member, IEEE) received the B.Sc. (with high Hons.) and  M.Sc. degrees in electrical and electronics engineering from Bogazici University, Istanbul, Turkiye, in 2021 and Koc University, Istanbul, Turkiye, in 2023, respectively. He is currently working towards the Ph.D. degree in electrical and electronics engineering at Koc University, Istanbul, Turkiye. He is also a researcher at HİSAR Lab., TUBITAK BILGEM. His research interests include reconfigurable intelligent surfaces and integrated sensing and communication.
\end{IEEEbiographynophoto}\vspace{-40pt}

\begin{IEEEbiographynophoto}{ALI FUAT SAHIN} 
received the dual B.Sc. degrees in Electronics and Communication Engineering and Control and Automation Engineering from Istanbul Technical University (ITU), Istanbul, Türkiye, in 2023, where he is currently pursuing the M.Sc. degree in Telecommunication Engineering. He is also a Researcher with the Communication and Signal Processing Research Laboratory, TÜBİTAK BİLGEM. His research interests include wireless communications, O-RAN architectures, reconfigurable intelligent surfaces (RIS), and explainable AI (XAI).
\end{IEEEbiographynophoto}\vspace{-40pt}

\begin{IEEEbiographynophoto}{ONUR SALAN} 
received the B.Sc. degree in Electronics and Communication Engineering from Izmir Institute of Technology, Izmir, Türkiye, in 2016, and the M.Sc. degree in Electronics and Communication Engineering from Yildiz Technical University, Istanbul, Türkiye, in 2023. He is currently pursuing the Ph.D. degree in Telecommunications Engineering at Istanbul Technical University, Istanbul, Türkiye. Since 2016, he has been a researcher with the Informatics and Information Security Research Center (TUBITAK BILGEM). His research interests include O-RAN architecture, artificial intelligence, and reconfigurable intelligent surfaces.
\end{IEEEbiographynophoto}\vspace{-20pt}

\begin{IEEEbiographynophoto}{RECEP A. TASCI}
(Graduate Student Member, IEEE) received the B.S. degree in electrical and electronics engineering from Istanbul Medipol University and the M.S. degree in electrical and electronics engineering from Koc University, Turkiye, in 2020 and 2022, respectively, where he is currently pursuing the Ph.D. degree. He is a Research Fellow at Medipol University, Turkiye. His research interests include wireless communications, reconfigurable intelligent surfaces, channel modeling, signal processing, zero-power and thermal noise communications, and KLJN key exchange.
\end{IEEEbiographynophoto}\vspace{-40pt}

\begin{IEEEbiographynophoto}{RECEP VURAL} 
received the B.S. degree in Electrical and Electronics Engineering from Istanbul Medipol University, Istanbul, Türkiye, in 2022, and the M.S. degree in Electrical and Electronics Engineering from Koç University, Istanbul, Türkiye, in 2025. He is currently working as a doctoral researcher at Tampere University, Tampere, Finland. His research interests include semantic communications, reconfigurable intelligent surfaces and physical layer security.

\end{IEEEbiographynophoto}\vspace{-40pt}

\begin{IEEEbiographynophoto}{YUSUF ISLAM TEK} 
received the B.S. degree in Electrical and Electronics Engineering from Nuh Naci Yazgan University, Kayseri, Türkiye, in 2020, and the M.S. degree in Electrical and Electronics Engineering from Koç University, Istanbul, Türkiye, in 2023, where he is currently pursuing the Ph.D. degree. Since 2025, he has been working as an R\&D Engineer with Türk Telekom, Ankara, Türkiye. His research interests include physical layer security, high-mobility waveform design, noise-driven communications, and artificial intelligence applications for next-generation wireless communication systems.
\end{IEEEbiographynophoto}\vspace{-40pt}

\begin{IEEEbiographynophoto}{ERTUGRUL BASAR} 
(Fellow, IEEE) received the B.S. degree from Istanbul University in 2007 and the M.Sc. and Ph.D. degrees from Istanbul Technical University in 2009 and 2013, respectively. He is currently Professor of Wireless Systems at Tampere University, Finland, and Head of the Wireless Systems Group at the Tampere Wireless Research Centre. His research interests include 6G and beyond wireless systems, reconfigurable intelligent surfaces, index modulation, MIMO communications, physical layer security, quantum communications, noise-driven communications, and AI-enabled wireless networks. He has authored/co-authored more than 300 publications in leading journals and conferences and has served the wireless communications community in various editorial and technical roles.
\end{IEEEbiographynophoto}\vspace{-40pt}

\begin{IEEEbiographynophoto}{IBRAHIM HOKELEK} 
(Member, IEEE) received the B.Sc. and M.Sc. degrees in electrical and electronics engineering from Bilkent University, Ankara, Türkiye, and the Ph.D. degree from the Graduate Center, The City University of New York (CUNY). He was a Senior Research Scientist with Telcordia Technologies Inc., Piscataway, NJ, USA, from 2006 to 2010. He has been the Project Manager and the Chief Researcher of the Scientific and Technological Research Council of Türkiye (TÜBİTAK). He has been an Adjunct Professor with the Electronics and Communication Engineering Department, Istanbul Technical University. His research interests include wireless mobile networks, network planning and management, and deterministic networks.
\end{IEEEbiographynophoto}\vspace{-40pt}

\begin{IEEEbiographynophoto}{ALI GORCIN} 
is the Head of TÜBİTAK BİLGEM Center. Since 2020, Assoc. Prof. Dr. GORCIN is also a faculty member at Istanbul Technical University, Department of Electronic Communication Engineering. He completed his undergraduate education at Istanbul Technical University, Department of Electronic Communication Engineering. Afterward, he started to work as a researcher at TÜBİTAK MAM BTAE, and during this period, he completed his master's degree at the same university.
Assoc. Prof. GORCIN, after working at TÜBİTAK for about 6 years, went to the University of South Florida in the United States of America for his doctoral studies and completed his studies in the field of wireless communication. After his doctoral studies, he worked in different positions in the private sector in the United States. In 2016, he returned to Turkey and started working at Yıldız Technical University and then was appointed as the Vice President of Test and Evaluation at TÜBİTAK BİLGEM.
\end{IEEEbiographynophoto}\vspace{-40pt}

\begin{IEEEbiographynophoto}{KARIM BOUTIBA} 
(Member, IEEE) received the Ph.D. degree from Sorbonne University in 2024. He is a Researcher with the Communication Systems Department, EURECOM. He is working toward upgrading 5G systems to 6G using EURECOM’s OpenAirInterface (OAI)-based test platform. He was involved in collaborative research projects and an Active Contributor to the OAI projects. His research interests include next-generation networking, 5G new radio, network slicing, open RAN, optimization algorithms, and reinforcement learning for 5G networks and beyond.
\end{IEEEbiographynophoto}\vspace{-40pt}

\begin{IEEEbiographynophoto}{ADLEN KSENTINI}
is a full professor in the Communication Systems Department at EURECOM, where he leads the Netsoft group, a team of 23 researchers dedicated to advancing network softwarization for 5G and 6G technologies. His research interests include network softwarization and cloudification, with a particular focus on applying machine learning (ML) and artificial intelligence (AI) to 5G and 6G networks. He has received several prestigious awards, including Best Paper Awards from IEEE Globecom 2025, IEEE IWCMC 2016, IEEE ICC 2012, and ACM MSWiM 2005. In 2017, he was honored with the IEEE ComSoc Fred W. Ellersick Award for the best paper published in IEEE Communications Magazine.
\end{IEEEbiographynophoto}\vspace{-20pt}

\end{document}